\documentclass[aps,prd,twocolumn,groupedaddress,showpacs]{revtex4-1}
\usepackage[pdftex]{graphicx}
\usepackage{subfigure}
\usepackage{sidecap}
\usepackage{multirow}
\usepackage{textcomp}

\usepackage{amsmath}
\usepackage{hyperref}
\usepackage{color}

\begin{document}

%\preprint{LIGO-P1400164}

\def\gw{gravitational-wave}

\title{Prospects for doubling the range of Advanced LIGO}

\author{John Miller}
\email{jmiller@ligo.mit.edu}
\author{Lisa Barsotti}
\author{Salvatore Vitale}
\author{Peter Fritschel}
\affiliation{LIGO Laboratory, Massachusetts Institute of Technology, 185 Albany St, Cambridge, MA 02139, USA}
\author{Daniel Sigg}
\affiliation{LIGO Hanford Observatory, PO Box 159, Richland, WA 99352, USA}
\author{Matthew Evans}
\affiliation{LIGO Laboratory, Massachusetts Institute of Technology, 185
  Albany St, Cambridge, MA 02139, USA}

\date{\today}

\begin{abstract}
  In the coming years, the gravitational-wave community will be
  optimising detector performance to target a variety of astrophysical
  sources which make competing demands on detector sensitivity in
  different frequency bands.  In this paper we describe a number of
  technologies that are being developed as anticipated upgrades to the
  Advanced LIGO detectors and quantify the potential sensitivity
  improvement they offer. Specifically, we consider squeezed light
  injection for the reduction of quantum noise, detector design and
  materials changes which mitigate thermal noise and mirrors with
  significantly increased mass.  We explore how each of these
  technologies impacts the detection of the most promising
  gravitational-wave sources and suggest an effective progression of
  upgrades which culminates in a twofold improvement in broadband
  sensitivity.
\end{abstract}

\pacs{04.80.Nn, 95.55.Ym, 95.85.Sz, 07.60.Ly}
%\keywords{}

\maketitle

%%%%%%%%%%%%%%%%%%
\section{Introduction}
A world-wide network of ground-based gravitational-wave
detectors~\cite{Deg12a, Fritschel2014, Aso13a} promises to begin the
era of gravitational-wave astronomy by detecting ripples in space-time
produced by astrophysical sources such as coalescing binary neutron
stars (BNS) and binary black holes (BBH). These kilometre-scale
Michelson-style interferometers are designed to detect
gravitational-wave strain amplitudes of $10^{-23}$ or smaller.

While Advanced LIGO and other advanced detectors are expected to
observe tens of compact binary coalescence (CBC) events per
year~\cite{RateDoc}, great benefit can be gained by further extending
their astrophysical reach~\cite{Sathyaprakash:2012jt,3G:Science}. This
is particularly true for Bayesian studies which combine evidence from
multiple detections. For example, it has been shown that
gravitational-wave signals recorded by Advanced LIGO can be used to
perform strong-field tests of General
Relativity~\cite{Li,LiAl,Agathos} and measure the equation of state of
neutron stars~\cite{DelPozzo,Wade, Wade1}. Improved sensitivity would
lead to more frequent detections, allowing more powerful tests to be
performed. A larger catalogue of CBC data would also facilitate study
of the distribution of neutron star and black hole masses. Such
information enables one to comment on the maximum possible neutron
star mass and probe the existence of a mass gap between black holes
and neutron stars~\cite{FarrW}.  Finally, more regular observation of
BNS events increases the probability of identifying an electromagnetic
counterpart, which may be used to, amongst other things, determine the
Hubble constant~\cite{2013arXiv1307.2638N} (although we note that
cosmological parameters can be estimated with gravitational-waves
alone~\cite{2012PhRvL.108i1101M,2012PhRvD..86d3011D}).

In addition to yielding more frequent detections, improving the
sensitivity of Advanced LIGO also increases the likelihood of
witnessing rare sources. For instance, the existence of intermediate
mass black holes ($M$$\sim$$\mathcal{O}(10^2-10^3)M_\odot$) is still
controversial\cite{Miller2003}. A single detection would provide the first direct proof
of their existence, whereas multiple detections could be used to
constraint their formation rate and test the no-hair
theorem~\cite{GossanAl}.

5- to 10-fold strain sensitivity improvements are possible but can
only be achieved by significantly modifying core components of the
existing interferometers~\cite{Rana:RMP}, by constructing a new
ultra-high-vacuum envelope to accommodate longer interferometer
arms~\cite{Lungo14} or both~\cite{ET}. In this paper we analyse a
progression of upgrades to Advanced LIGO which do not require changes
to the current buildings or vacuum infrastructure and, as much as
possible, leverage proven technologies. Whilst such upgrades have
previously been discussed in isolation we, for the first time, examine
realistic combinations of these upgrades and provide a coherent
strategy and timeline for their implementation.  We find that, in
unison, these upgrades achieve a factor of two broadband sensitivity
improvement. This translates into a detection rate increase
approaching one order of magnitude.

%%%%%%%%%%%%%%%%%%
\section{Sensitivity improvement targets}

\begin{figure}[t!]
\centering
\includegraphics[width=1\columnwidth]{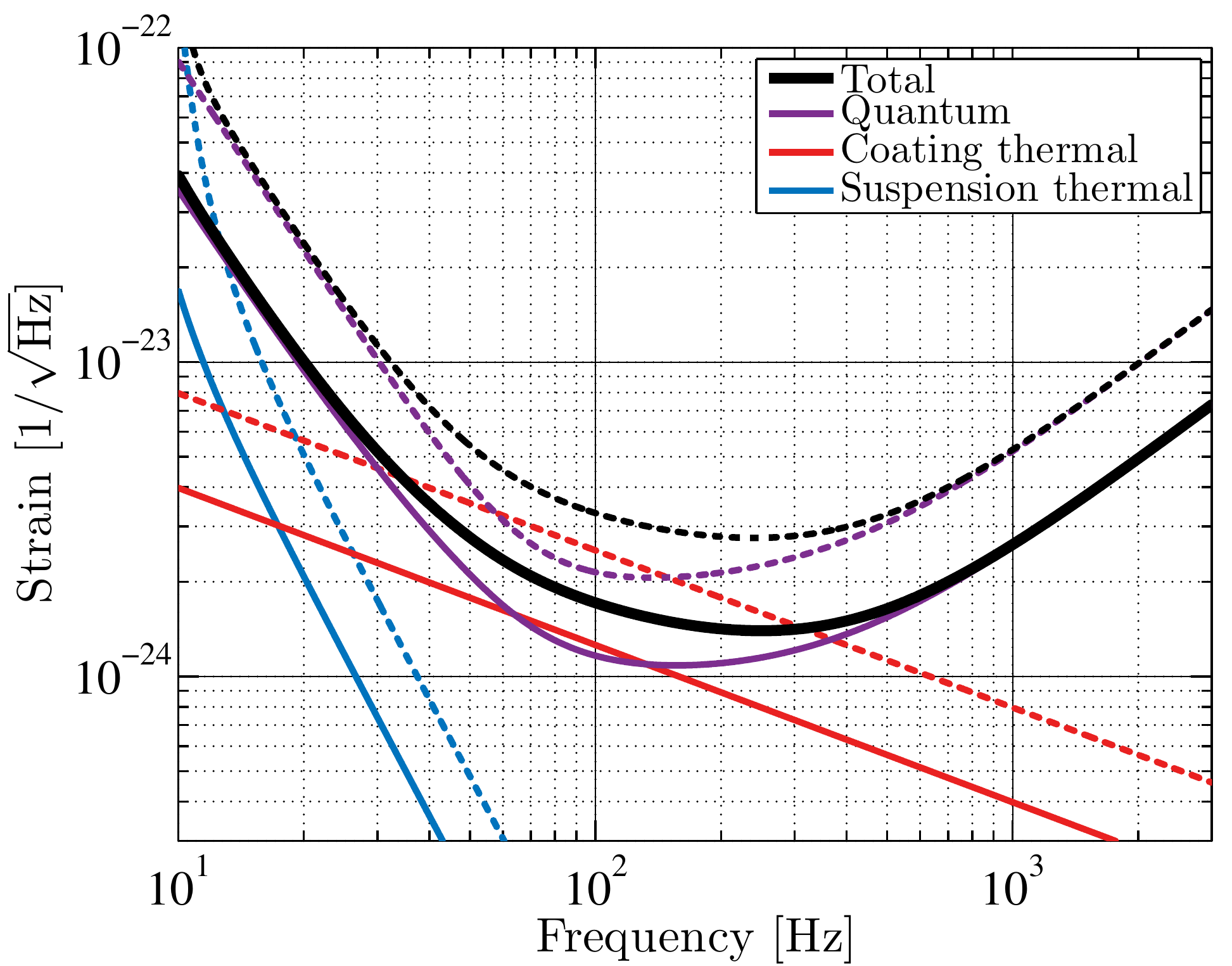}
\caption{Strain sensitivity of a possible upgraded Advanced LIGO
  interferometer. Improved thermal noise (factor of two), improved
  quantum noise (16\,m filter cavity and 6\,dB of measured squeezing
  at high frequency) and heavier test masses (also a factor of two)
  are assumed. The equivalent Advanced LIGO curves are shown using
  dashed lines. \label{fig:fig1}}
\end{figure}

In this work we focus on noise sources driven by fundamental physical
processes, acknowledging that, in common with all previous generations
of interferometer, technical and non-Gaussian noise must be mitigated
before fundamental noise sources are exposed.

%independent of interferometer configuration,

Quantum noise and thermal noise are the principal fundamental noise
sources in Advanced LIGO (see Fig.~\ref{fig:fig1}). Quantum noise is
produced by shot noise, the statistical fluctuations in the arrival
time of photons at the interferometer output, and by radiation
pressure noise, the fluctuations in the number of photons impinging on
the test masses~\cite{Cav80a, Cav81a}.

Thermal noise may be explained by the fluctuation-dissipation theorem,
which connects displacement fluctuations in a system surrounded by a
thermal bath with dissipation caused by internal
losses~\cite{Levin08}. The dominant sources of thermal noise in
current interferometric gravitational-wave detectors are the
high-reflectivity optical coatings on the test masses (coating thermal
noise) and the fused-silica fibres which suspend the test masses
against the force of Earth's gravity (suspension thermal
noise)~\cite{Saulson1990}.

Squeezed light injection is a proven technique to reduce quantum
noise. New lower-loss coating materials or the current optimised
amorphous coatings in combination with a larger beam size on the test
masses could reduce coating thermal noise by a factor of two or
more. Two other possible improvements are considered here: the
reduction of suspension thermal noise, which contributes to the total
noise below 50\,Hz, and a set of heavier test masses, which reduces
the impact of radiation pressure noise in the same frequency band.

\begin{table}[b!]
\renewcommand{\arraystretch}{0.8}
\caption{\label{tab:tab1}%
  BNS and BBH ranges for an Advanced LIGO interferometer in which combinations of the main limiting noise sources have been reduced in the manner described in the text. A plausible incremental progression of upgrades is highlighted in blue: (i) quantum noise reduction through squeezed light injection, (ii) a factor of two reduction in coating thermal noise, (iii) a factor of two increase in the mirror mass and (iv) a factor of two reduction in suspension thermal noise.}
\begin{ruledtabular}
    \begin{tabular}{p{0.01cm}cccc|cc}
    &\multicolumn{4}{c|}{Improved quantity} &       \multicolumn{2}{c}{Range}\\ \hline
      &\multirow{2}{*}{Quantum} & Coating & Mirror & Suspension & BNS& BBH\\
&&thermal&mass&thermal&[Mpc]&[Gpc]\\
\colrule
&---                         & ---                         & ---                         & ---                         & 220                   & 1.3                  \\
\textcolor{blue}{(i)}&\textcolor{blue}{$\bullet$} & \textcolor{blue}{---}       & \textcolor{blue}{---}       & \textcolor{blue}{---}       & \textcolor{blue}{280} & \textcolor{blue}{1.5}\\
&---                         & $\bullet$                   & ---                         & ---                         & 280                   & 1.7                  \\
&---                         & ---                         & $\bullet$                   & ---                         & 260                   & 1.6                  \\
&---                         & ---                         & ---                         & $\bullet$                   & 220                   & 1.3                  \\
\textcolor{blue}{(ii)}&\textcolor{blue}{$\bullet$} & \textcolor{blue}{$\bullet$} & \textcolor{blue}{---}       & \textcolor{blue}{---}       & \textcolor{blue}{400} & \textcolor{blue}{2.3}\\
&$\bullet$                   & ---                         & $\bullet$                   & ---                         & 320                   & 1.9                  \\
&$\bullet$                   & ---                         & ---                         & $\bullet$                   & 280                   & 1.6                  \\
&---                         & $\bullet$                   & $\bullet$                   & ---                         & 350                   & 2.3                  \\
&---                         & $\bullet$                   & ---                         & $\bullet$                   & 280                   & 1.7                  \\
&---                         & ---                         & $\bullet$                   & $\bullet$                   & 270                   & 1.7                  \\
\textcolor{blue}{(iii)}&\textcolor{blue}{$\bullet$} & \textcolor{blue}{$\bullet$} & \textcolor{blue}{$\bullet$} & \textcolor{blue}{---}       & \textcolor{blue}{470} & \textcolor{blue}{2.9}\\
&$\bullet$                   & $\bullet$                   & ---                         & $\bullet$                   & 410                   & 2.3                  \\
&$\bullet$                   &                             & $\bullet$                   & $\bullet$                   & 320                   & 1.9                  \\
&---                         & $\bullet$                   & $\bullet$                   & $\bullet$                   & 350                   & 2.3                  \\
\textcolor{blue}{(iv)}&\textcolor{blue}{$\bullet$} & \textcolor{blue}{$\bullet$} & \textcolor{blue}{$\bullet$} & \textcolor{blue}{$\bullet$} & \textcolor{blue}{480} & \textcolor{blue}{3.0}
    \end{tabular}
\end{ruledtabular}
\end{table}

Fig.~\ref{fig:fig1} shows the resulting quantum noise and thermal
noise after implementation of all of the improvements discussed in the
previous paragraph.

The canonical figure of merit describing the sensitivity of a detector
is its range. In this work we define the average BNS range to be
$\sqrt[3]{3/64\times1.8375}\simeq1/2.26$ times the redshift-corrected
luminosity distance at which an optimally oriented and located BNS
system consisting of two 1.4\,$M_{\odot}$ neutron stars would give a
matched-filter signal-to-noise ratio of 8 in a single
detector~\cite{Finn1993,Finn1996,Allen2012}. We also consider the
analogous BBH range for a system consisting of two 10\,$M_{\odot}$
black holes. The baseline Advanced LIGO interferometers should each
achieve a BNS range of up to 220\,Mpc and a BBH range of up to
1.3\,Gpc.

Table~\ref{tab:tab1} shows how subsets of the improvements we consider
affect the detector's range. In isolation, none of these improvements
guarantees more than a 30\% increase in range with respect to Advanced
LIGO. However, once combined, the range can be approximately
doubled. The status and prospects of these technologies are described
in the following sections.

%%%%%%%%%%%%%%%%%%
\section {Squeezed light for quantum noise reduction}

\begin{figure}[t!]
\centering
\includegraphics[width=1\columnwidth]{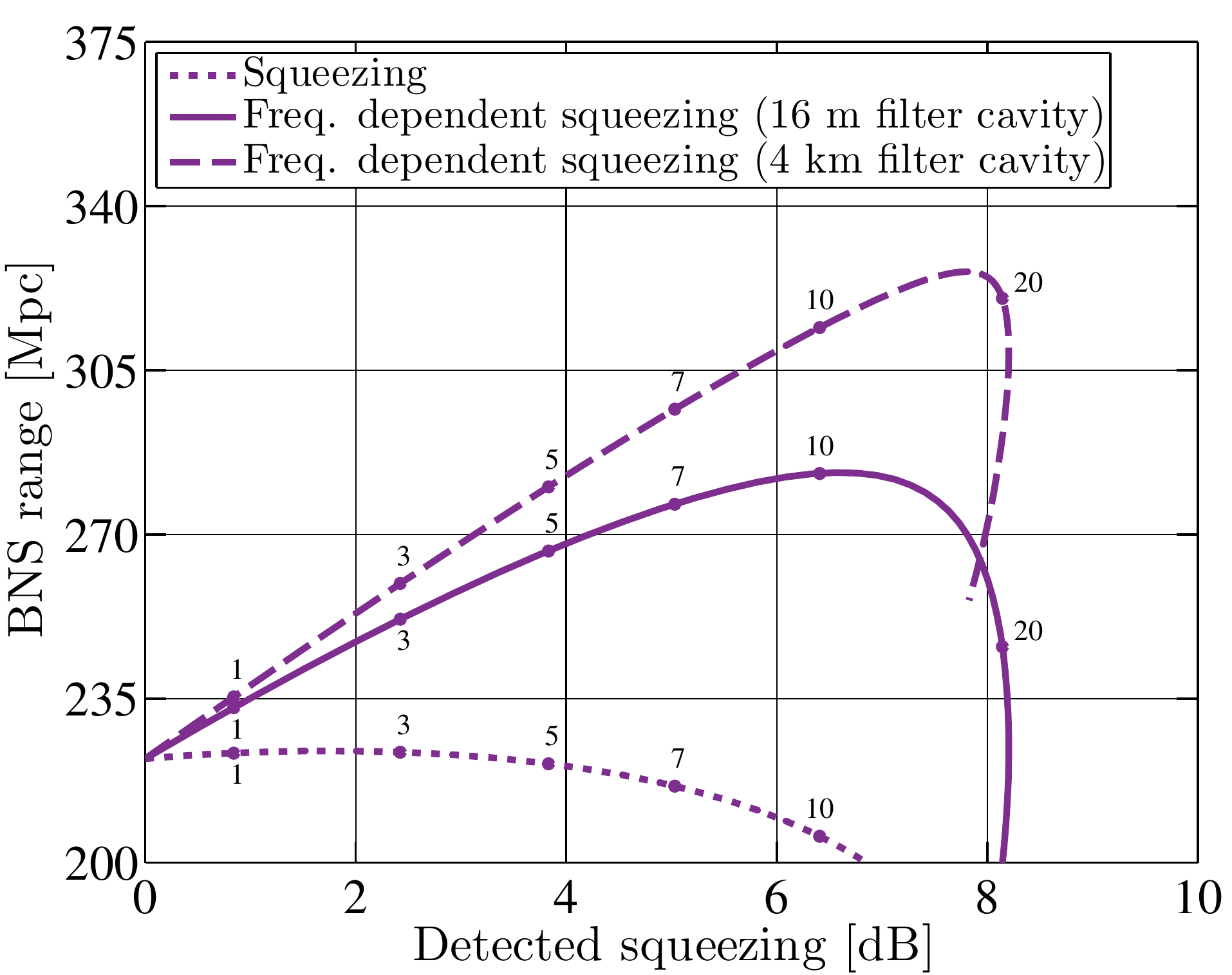}
\caption{Impact of squeezing in Advanced LIGO on BNS range as a
  function of measured high-frequency quantum noise
  reduction. Numerical labels indicate the magnitude of squeezing
  injected in dB. The best experimentally detected level currently
  stands at approximately 12 dB \cite{Stefszky12}. A longer filter
  cavity does not significantly improve the effectiveness of squeezing
  in Advanced LIGO.\label{fig:quantum1}}
\end{figure}

Squeezed states of light~\cite{Chua14} have already been employed to
improve the sensitivity of gravitational-wave
interferometers~\cite{LIG11b, Bar13a} and are routinely used in
GEO600~\cite{Grote2013}. However, any reduction in quantum shot noise
at high frequencies is accompanied by a commensurate increase in
quantum radiation pressure noise. If applied to Advanced LIGO,
squeezing would reshape the sensitivity of the detector as a function
of frequency to the detriment of BNS range (see
Fig.~\ref{fig:quantum1}).

% A number of loss
% mechanisms couple anti-squeezing into the measurement
% quadrature~\cite{Kwee14a}. These effects are most noticeable with
% high levels of injected squeezing and are responsible for the
% curious double-valued nature of the frequency dependent squeezing
% curves.

By reflecting a squeezed state from a detuned high-finesse optical
resonator, known as a filter cavity, one can produce frequency
dependent squeezing which simultaneously reduces shot noise at high
frequencies and radiation pressure noise at low
frequencies~\cite{Kim01a, Har03a}.

Interferometers measure the projection of the quantum noise ellipse
onto the gravitational-wave signal (see e.g.~Figure 5
of~\cite{Kim01a}). Technical effects can cause the relative
orientation of the ellipse to oscillate as a function of time (phase
noise) or introduce noise from the orthogonal axis of the ellipse
(mode matching errors, filter cavity loss), coupling anti-squeezing
into the measurement quadrature and reducing the level of observed
squeezing~\cite{Kwee14a}. Increasing the level of squeezing increases
the eccentricity of the quantum noise ellipse, making one more
susceptible to technical noise of this kind. Thus, more squeezing can
lead to reduced sensitivity, as evidenced by the curious double-valued
nature of the frequency dependent squeezing curves in
Fig.~\ref{fig:quantum1}.

With a realistic implementation of frequency dependent squeezing (see
parameters in Table~\ref{tab:param} and methods described
in~\cite{Kwee14a}), broadband improvements are available, leading to
increases of 30\% and 15\% in BNS and BBH ranges, respectively, and
valuable improvements in our ability to extract gravitational-wave
signal parameters for astrophysical investigations~\cite{Lynch2014}.

With 10\,dB of injected squeezing, realistic loss mechanisms, mainly
the filter cavity intra-cavity loss, limit the range achievable with a
16~m filter cavity to 280\,Mpc.  A longer filter cavity can mitigate
the impact of intra-cavity losses but, as shown in
Fig.~\ref{fig:quantum1}, would only yield a 10\% improvement. For this
reason, a 16-20\,m filter cavity is the baseline option for upgrading
Advanced LIGO~\cite{Eva13a}. To date, frequency dependent squeezing
has only been demonstrated at radio frequencies~\cite{Che05a}. The
research to transfer this technique to gravitational-wave frequencies
is ongoing~\cite{Isogai13, Kwee14a}.

\begin{table}
\renewcommand{\arraystretch}{0.8}
  \caption{\label{tab:param}Parameters used in evaluating the
    performance of an Advanced LIGO interferometer incorporating frequency-dependent squeezing. Interferometer parameters are as given in Table~I of \cite{Kwee14a}. Values in parentheses correspond to an interferometer with 80\,kg mirrors.}
\begin{ruledtabular}
%\begin{tabular}{p{5.25cm}r}
\begin{tabular}{lr}
Parameter                                 & Value                  \\
\colrule
Filter cavity length                      & 16\,m                  \\
Filter cavity input mirror transmissivity & 67(47)\,ppm            \\
Filter cavity detuning                    & 49(35)\,Hz             \\
Filter cavity half-bandwidth              & 56(41)\,Hz             \\
% Filter cavity input                     &
% \multirow{2}{*}{66.3~ppm}                                        \\
% mirror transmissivity                   &                        \\
Filter cavity losses                      & 8\,ppm                 \\
Injection losses                          & $5\%$                  \\
Readout losses                            & $5\%$                  \\
Mode-mismatch (squeezer-filter cavity)    & $2\%$                  \\
Mode-mismatch (squeezer-interferometer)   & $5\%$                  \\
% Mode-mismatch                           & \multirow{2}{*}{$2\%$} \\
% (squeezer-filter cavity)                &                        \\
% Mode-mismatch                           & \multirow{2}{*}{$5\%$} \\
% (squeezer-local oscillator)             &                        \\
Frequency independent phase noise (RMS)   & 5\,mrad                \\
Filter cavity length noise (RMS)          & 0.3\,pm                \\
Injected squeezing                        & 9.1\,dB
\end{tabular}
\end{ruledtabular}

\end{table}

%%%%%%%%%%%%%%%%%%
\section{Increased mirror mass}

Radiation pressure noise scales inversely with the core optics' mass,
providing a means of mitigating its impact. The test masses in
Advanced LIGO are made from ultra-low absorption fused silica. This
material is available in sizes allowing for up to about twice the
present mass. We thus establish 80\,kg as the fiducial mass of our
upgraded mirrors. Such an increase demands new suspensions capable of
supporting a greater load. However, this can be achieved by mimicking
the current suspension design, with straightforward enlargement of
springs, fibres and mass elements. The existing seismic isolation
system can accommodate the increased mass without modification.

Larger masses should allow the use of larger beams, leading to lower
coating thermal noise. From this standpoint, the optimal mirror aspect
ratio (radius/thickness) is approximately unity. With such a geometry,
coating thermal noise amplitude scales as $m^{-1/3}$, where $m$ is the
mirror mass, if the diffraction loss is held constant. Increased mass
also reduces suspension thermal noise. This effect scales as
$m^{-1/4}$ in amplitude \cite{Saulson1990} and has been included in
the values presented in this work.

%%%%%%%%%%%%%%%%%%
\section {Coating thermal noise reduction}

The optical coatings used in gravitational-wave detectors have
extraordinary optical properties: absorption below 1\,ppm, scatter
losses around 10\,ppm and very tightly controlled
reflectivities. Conversely, these coatings are mechanically much more
lossy than the mirror substrates on which they are deposited. Thus,
they constitute the dominant source of thermal noise~\cite{Harry02}.

Coating research has received considerable attention in the past
decade as the use of resonant optical cavities has become widespread
in frequency standards, gravitational-wave detectors and other
precision optical measurements~\cite{Harry2012}. Informed by this
work, the coatings used in Advanced LIGO are composed of alternating
layers of amorphous silica and titania-doped tantalum
pentoxide~\cite{Harry07,Evans08}. Despite ongoing research, amorphous
materials offering lower losses while maintaining acceptable optical
properties have proven elusive~\cite{Flaminio2010}.  This has led to a
search for new coating materials and technologies for use in future
gravitational-wave detectors.

One potential solution is the use of crystalline coatings.  A leading
candidate is epitaxial layers of Al$_x$Ga$_{1-x}$As, where the
parameter $x$ takes two values to provide the high and low refractive
index materials of a multilayer Bragg reflector. Such mirrors can be
grown on a GaAs wafer and then transferred to a fused-silica substrate
and have been shown to provide at least a factor of three reduction in
the amplitude of coating thermal noise~\cite{Cole2013}. While
crystalline coatings are promising, they have yet to be demonstrated
on a 50\,cm-scale mirror. Scaling up this technology presents several
technical challenges, both in the manufacturing process and in meeting
the extremely stringent surface-figure specifications associated with
multi-kilometre resonant cavities.

%%%%%%%%%%%%%%%%%%
\section{Larger Beams}

Increasing the size of the laser beam reflected from the test masses
reduces the impact of coating thermal noise in proportion to the beam
diameter, simply due to averaging over a larger coating
area~\cite{Harry2012}. While this solution is conceptually simple,
feasible beam diameters are limited by optic size, optical stability
considerations in the arm cavities and challenges in fabricating
suitable mirror surfaces. In the context of Advanced LIGO and current
coating technology, even with larger mirrors, no more than a factor of
two reduction in coating thermal noise is likely to be achieved
without compromising the performance of the interferometer---either
due to clipping losses on the optics or angular instabilities in the
interferometer~\cite{SidlesSigg, Dooley2013}.

Given the potential difficulties associated with these approaches it
is not certain that either will achieve its theoretically predicted
performance on the appropriate timescale ($\sim$5 years). We therefore
expect that a combination of larger beams and new coating materials
will be implemented in order to reduce coating thermal noise and
reduce our expectation for the available improvement from a factor of
approximately six to just a factor of two.

%%%%%%%%%%%%%%%%%%
\section{Suspension thermal noise reduction}

As part of a multi-stage seismic isolation system, the test masses of
the Advanced LIGO interferometers are suspended from four 60\,cm-long
low-loss fused silica fibres. The fibres have a circular cross section
whose diameter varies to best cancel thermoelastic damping, to
maintain high bounce and violin mode frequencies and to ease handling
and bonding to the test masses. Thermal noise from this suspension
system dominates the total thermal noise below $\sim$20\,Hz (see
Fig.~\ref{fig:fig1}). Fortunately, several low-risk methods are
available for reducing suspension thermal noise~\cite{Hammond2012}.

The amplitude of suspension thermal noise scales as $1/l$, where $l$
is the length of the suspension fibres. This scaling includes equal
contributions from ``dissipation dilution'' and the improved isolation
resulting from a downward shift of the resonant
frequency~\footnote{Increasing the suspension length brings the added
  benefit of reducing thermal noise in the vertical direction, which
  couples to the longitudinal direction due to the curvature of the
  Earth.}. Further gains can be realised by refining the geometry of
the fibre ends to improve dilution factors and by heat treatment of
the fibres to reduce surface losses. One recent study estimates that
the above techniques can reduce the amplitude of suspension thermal
noise by a factor of 2.5~\cite{Hammond2012}. In this work we
conservatively assume a factor of two, realising this improvement
solely through increased suspension length and neglecting violin
modes.

%%%%%%%%%%%%%%%%%%%%%%%%%
\section{Discussion}

Each of these potential improvements to Advanced LIGO can offer a
factor of two reduction in the corresponding noise term. However, each
noise term contributes significantly only at certain frequencies (see
Fig.~\ref{fig:fig1}) and none is sufficiently dominant to effect more
than a 30\% change in BNS or BBH range (see Table~\ref{tab:tab1}). A
significant increase in inspiral range is only realised when
improvements are implemented simultaneously.

\begin{figure}[t!]
\centering
\includegraphics[width=\columnwidth]{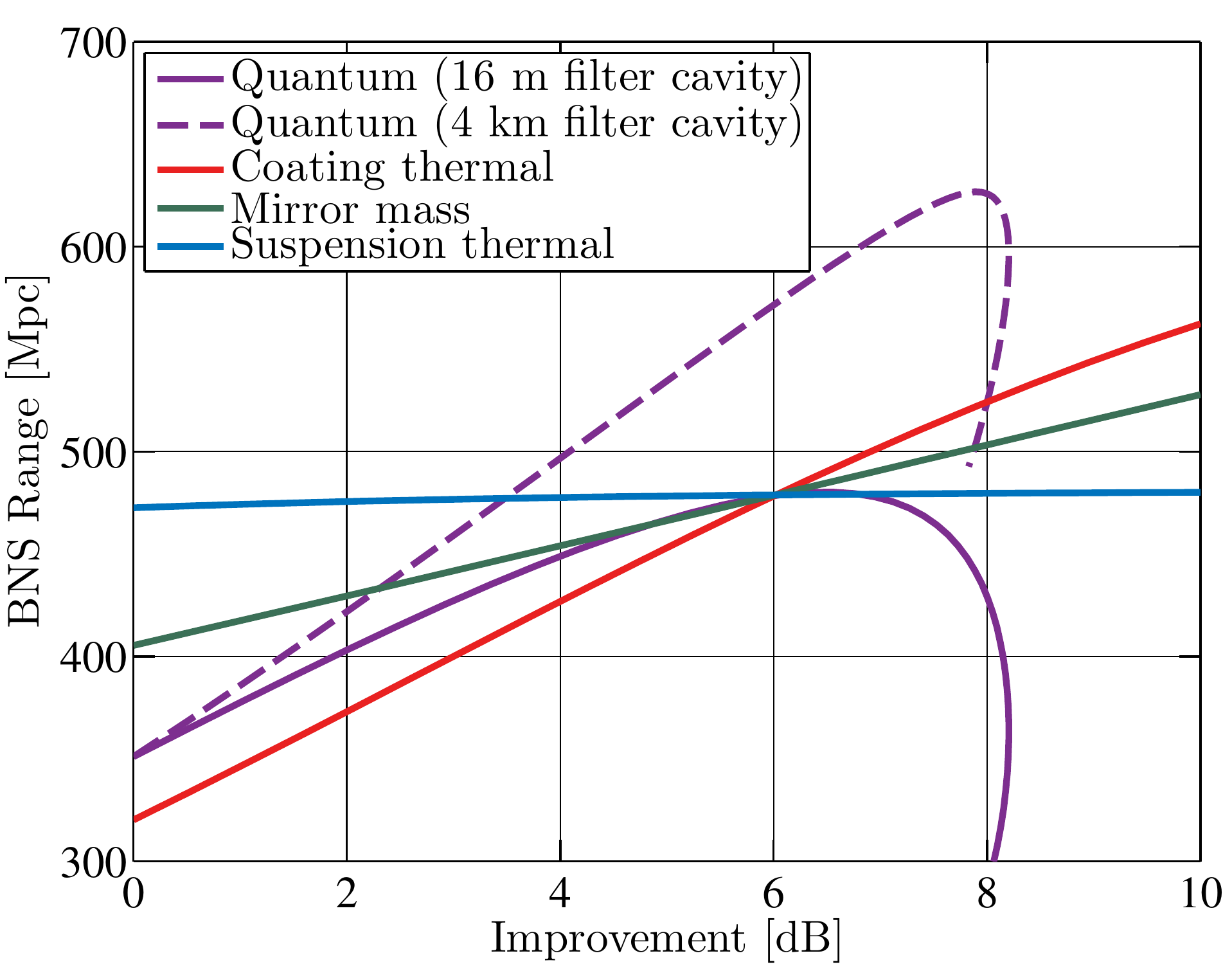}
\caption{BNS range as a function of improvement in the given
  quantities assuming all other upgrades have been implemented. For
  quantum noise, `Improvement' describes the measured level of
  high-frequency squeezing, for mirror mass it expresses the mass
  increase factor and for thermal terms it captures the noise
  reduction factor. The four continuous curves intersect at the 6\,dB
  point, where all noise terms have been reduced by a factor of two.
  In contrast to the baseline Advanced LIGO case, a longer filter
  cavity (dashed curve) proves significantly more effective in an
  upgraded interferometer.\label{fig:fig4}}
\end{figure}

Fig.~\ref{fig:fig4} conveys the relative importance of each potential
upgrade---{\it assuming all other improvements have already been
  made}. For example, the quantum noise curves show BNS range as
function of measured high-frequency squeezing for an interferometer in
which coating thermal noise and suspension thermal noise have been
reduced by a factor of two (or $\sim$6\,dB) and the mass of the
mirrors has been increased by a factor of two (all with respect to
Advanced LIGO). Coating thermal noise reductions are shown to have the
greatest effect on BNS range. Quantum noise improvements of up to 6~dB
have comparable impact. Beyond this point BNS range does not increase
monotonically with the level of detected squeezing due to currently
achievable levels of technical noise, as in Advanced LIGO (see
Fig.~\ref{fig:quantum1}). However, in contrast, a 4\,km long filter
cavity is significantly more beneficial in an upgraded interferometer,
since radiation pressure has become more influential at low
frequencies. The maximum achievable BNS range is 20\% higher with
respect to a short filter cavity---bringing the maximum range to
580\,Mpc for 6\,dB of detected high-frequency squeezing.

Mitigating suspension thermal noise offers relatively modest gains in
astrophysical output compared to the other approaches.  By extension,
this indicates that improvements in noise sources such as seismic
noise and gravitational gradient noise, which are also prominent at
very low frequencies, but less so than suspension thermal noise, will
be even less effective.

For the BNS and BBH systems used to define our figures of merit,
sensitivity around 100\,Hz is of utmost importance. However, the
astrophysical impact of Advanced LIGO will likely not be limited to
the detection of stellar-mass compact binaries. Indeed, several other
sources are predicted to emit gravitational radiation detectable by
ground-based interferometers. However, sensitivity to such systems may
not be improved by all of the modifications discussed above. For
example, many interesting sources emit gravitational waves in the
300-3000\,Hz band; including core-collapse supernovae~\cite{Ott} and
rapidly rotating neutron stars~\cite{ContinuousW} (and references
therein). In this frequency range, detector performance is limited
entirely by quantum shot noise and is thus only improved by squeezed
light injection.

% TIMESCALE%
%%%%%%%%%%%%%%%%%%%%%%
It is currently anticipated that Advanced LIGO will undertake its
first observing run (O1) in 2015, with subsequent runs of increasing
sensitivity and duration in 2016-2017 (O2) and 2017-2018 (O3). Full
design sensitivity should be achieved by 2019 \cite{Fritschel2014}. We
envisage that a squeezed light source will be installed after O2, with
test mass and coating thermal noise upgrades being implemented
following O3.

\section{Conclusions}

Several improvements to Advanced LIGO will be possible over the next
decade. Individually, these upgrades can be used to target particular
frequency bands. In this work we have selected the most likely
upgrades and, for the first time, evaluated how combinations of them
perform. When implemented together, we find that the upgrades
discussed herein offer a twofold increase in broadband sensitivity,
enriching our astrophysical understanding and enhancing the importance
of Advanced LIGO as a tool for multi-messenger astronomy.  The
progression of upgrades to Advanced LIGO involving squeezed light,
coating thermal noise reduction and heavier test masses offers a path
for improvement which will extend the volume of the observable
gravitational-wave universe by nearly an order of magnitude. While
expected detection rates are currently very uncertain, even in the
most pessimistic scenario these upgrades will take Advanced LIGO from
observing a few events per year to observing a few events per month, a
critical improvement as the era of gravitational-wave astronomy
begins.

%%%%%%%%%%%%%%%%%%%%%%
\begin{acknowledgments}
  The authors gratefully acknowledge the guidance of Rainer Weiss,
  Nergis Mavalvala, Rana Adhikari and David McClelland. They also
  thank Stefan Ballmer, Stefan Hild, Sheila Rowan and the other
  members of the LIGO Scientific Collaboration's Advanced
  Interferometer Configurations working group for useful
  discussions. LIGO was constructed by the California Institute of
  Technology and Massachusetts Institute of Technology with funding
  from the National Science Foundation and operates under cooperative
  agreement PHY-0757058. Advanced LIGO was built under award
  PHY-0823459. This paper has been assigned LIGO Document
  No. LIGO-P1400164.
\end{acknowledgments}

%%%%%%%%%%%%%%%%%%%%%%
\bibliography{bibliographyFile}

\end{document}